\documentclass[10pt,conference]{IEEEtran}

\usepackage[utf8]{inputenc} 
\usepackage[T1]{fontenc}    
\usepackage{hyperref}       
\usepackage{url}            
\usepackage{booktabs}       
\usepackage{amsfonts}       
\usepackage{nicefrac}       
\usepackage{microtype}      
\usepackage{xcolor}         
\usepackage{soul}
\soulregister\cite7
\soulregister\ref7
\soulregister\xinyue7
\soulregister\textsubscript{1}
\sethlcolor{white}
\newcommand{\hlmeta}[1]{\sethlcolor{white}\hl{#1}\sethlcolor{white}}
\usepackage{amsthm}
\usepackage{amsmath}
\usepackage{graphicx}
\usepackage{enumitem}
\usepackage{booktabs, listings, amssymb, xcolor}
\definecolor{darkgreen}{rgb}{0.0,0.5,0.0}
\definecolor{darkred}{rgb}{0.6,0.0,0.0}
\usepackage{array, makecell}
\newcolumntype{L}[1]{>{\raggedright\arraybackslash}p{#1}}
\usepackage{multirow}
\usepackage{letltxmacro}
\usepackage{adjustbox}
\usepackage{etoolbox}
\usepackage{algorithmic}
\usepackage{caption}
\usepackage{subcaption}
\usepackage{listings}
\IEEEoverridecommandlockouts

\makeatletter
\patchcmd{\verb}{\@noligs}{\small\@noligs}{}{}
\makeatother

\usepackage[ most ]{tcolorbox}
\tcbset{
  nllisting/.style={
    listing only,
    listing options={
      basicstyle=\small\ttfamily,   
      keepspaces=true,
      columns=fullflexible,
      breaklines=true,
    },
    colback=gray!5,        
    colframe=gray!70,      
    boxrule=0.4pt,         
    arc=1mm,               
    left=1.5mm, right=1.5mm, top=1mm, bottom=1mm,
  }
}

\LetLtxMacro{\oldtexttt}{\texttt}
\renewcommand{\texttt}[1]{{\small\oldtexttt{#1}}}

\usepackage{amsthm}
\theoremstyle{definition}
\newtheorem{definition}{Definition}

\theoremstyle{plain}
\newtheorem{theorem}{Theorem}
\newtheorem{lemma}{Lemma}

\newtheorem{corollary}{Corollary}

\usepackage{cascadia-code}

\definecolor{backbeige}{RGB}{242, 242, 234}
\definecolor{backlightgray}{RGB}{247, 247, 247}
\definecolor{backlightblue}{RGB}{230, 243, 255}
\definecolor{codelightblue}{RGB}{0, 0, 255}
\definecolor{codedarkblue}{RGB}{0, 0, 139}
\definecolor{codedarkerblue}{RGB}{25, 25, 112}
\definecolor{codered}{RGB}{255, 0, 0}
\definecolor{codegreen}{RGB}{58, 128, 0}
\definecolor{codegray}{RGB}{160, 160, 160}
\definecolor{codedarkgray}{RGB}{89, 96, 105}
\definecolor{codebrown}{RGB}{137, 50, 9}
\definecolor{codedarkred}{RGB}{199, 34, 41}
\definecolor{codedeepblue}{RGB}{14, 71, 162}
\definecolor{codeviolet}{RGB}{91, 51, 175}

\definecolor{vsblack}{cmyk}{0, 0, 0, 1}             
\definecolor{vswhite}{cmyk}{0, 0, 0, 0}
\definecolor{vsdarkgrey}{cmyk}{0, 0, 0, 0.97}      
\definecolor{vsgreen}{cmyk}{0.6, 0, 0.84, 0}       
\definecolor{vsblue}{cmyk}{0.65, 0.33, 0, 0.05}       
\definecolor{vsrose}{cmyk}{0, 0.26, 0.38, 0}       
\definecolor{vslavender}{cmyk}{0, 0.42, 0, 0.1}   
\definecolor{vslightblue}{cmyk}{0.35, 0, 0, 0}     
\definecolor{vsaqua}{cmyk}{0.65, 0, 0.23, 0}       
\definecolor{vslightgreen}{cmyk}{0.12, 0, 0.33, 0.05} 
\definecolor{vslightergreen}{RGB}{173, 201, 159}   
\definecolor{vsyellow}{RGB}{247, 216, 43}          
\definecolor{vslighteryellow}{RGB}{217, 217, 172}

\lstdefinestyle{cspsharp}{
    alsoletter={0123456789+-*/@},
    backgroundcolor=\color{backbeige},
    basicstyle=\scriptsize\ttfamily,
    breakatwhitespace=false,
    breaklines=true,
    commentstyle=\color{codegreen},
    columns=flexible,
    escapeinside={*@}{@*},
    frame=lr,
    framesep=5pt,
    framerule=0pt,
    keepspaces=true,
    keywordstyle=[1]\color{codelightblue},
    keywordstyle=[2]\color{codedarkblue},
    keywordstyle=[3]\color{codered},
    keywordstyle=[4]\bfseries\color{codedarkerblue},
    literate=*
        {-}{{\color{codered}-}}1
        {\%}{{\color{codered}\%}}1
        {>}{{\color{codered}>}}1
        {<}{{\color{codered}<}}1
        {=}{{\color{codered}=}}1
        {[}{{\color{codered}[}}1
        {]}{{\color{codered}]}}1
        {\&}{{\color{codered}\&}}1
        {|}{{\color{codered}|}}1
        {!}{{\color{codered}!}}1
        {?}{{\color{codered}?}}1
        {\#}{{\color{codered}\#}}1,
    morecomment=[l]{//},
    morecomment=[s]{/*}{*/},
    moredelim=[is][\color{codedarkerblue}\bfseries]{`}{`},
    morekeywords=[1]{assert, atomic, channel, deadlockfree, define, divergencefree, else, enum, false, hvar, if, ifa, ifb, include, nonterminating, reaches, refines, true, var, while, alphabet, import, deadline, timeout, interrupt, tau, new},
    morekeywords=[2]{0, 1, 2, 3, 4, 5, 6, 7, 8, 9, 10, 100, @},
    morekeywords=[3]{+, -, *, /},
    morekeywords=[4]{Skip, Stop},
    numbers=left,
    numbersep=10pt,
    numberstyle=\color{gray}\tiny,
    string=[b]{"},
    stringstyle=\color{codegreen},
    showspaces=false,                
    showstringspaces=false,
    showtabs=false,                  
    tabsize=2,
    xleftmargin=0.45cm
}

\lstdefinelanguage{json}{
  basicstyle=\ttfamily\small,
  numbers=left,
  numberstyle=\tiny,
  stepnumber=1,
  numbersep=5pt,
  showstringspaces=false,
  breaklines=true,
  frame=single,
  columns=fullflexible,
  morestring=[b]",
  literate=
    *{0}{{{\color{black}0}}}{1}
     {1}{{{\color{black}1}}}{1}
     {2}{{{\color{black}2}}}{1}
     {3}{{{\color{black}3}}}{1}
     {4}{{{\color{black}4}}}{1}
     {5}{{{\color{black}5}}}{1}
     {6}{{{\color{black}6}}}{1}
     {7}{{{\color{black}7}}}{1}
     {8}{{{\color{black}8}}}{1}
     {9}{{{\color{black}9}}}{1}
     {:}{{{\color{black}{:}}}}{1}
     {,}{{{\color{black}{,}}}}{1}
     {\{}{{{\color{black}{\{}}}}{1}
     {\}}{{{\color{black}{\}}}}}{1}
     {[}{{{\color{black}{[}}}}{1}
     {]}{{{\color{black}{]}}}}{1}
}

\lstnewenvironment{cspsharp}[0]{
    \lstset{style=cspsharp}
} {}

\newcommand{\xinyue}[1]{{\color{red}X: #1}}

\title{PAT-Agent: Autoformalization for Model Checking}

\author{\IEEEauthorblockN{Xinyue Zuo}
\IEEEauthorblockA{\textit{National University of Singapore}\\
zuoxy@nus.edu.sg}
\and
\IEEEauthorblockN{Yifan Zhang}
\IEEEauthorblockA{\textit{National University of Singapore}\\
yifan.zhang\_@u.nus.edu}
\and
\IEEEauthorblockN{Hongshu Wang}
\IEEEauthorblockA{\textit{National University of Singapore}\\
hongshu.wang@u.nus.edu}
\and
\IEEEauthorblockN{Yufan Cai\IEEEauthorrefmark{1}}
\IEEEauthorblockA{\textit{National University of Singapore}\\
caiyf@nus.edu.sg}
\and
\IEEEauthorblockN{Zhe Hou}
\IEEEauthorblockA{\textit{Griffith University}\\
z.hou@griffith.edu.au}
\and
\IEEEauthorblockN{Jing Sun}
\IEEEauthorblockA{\textit{University of Auckland}\\
jing.sun@auckland.ac.nz}
\and
\IEEEauthorblockN{Jin Song Dong}
\IEEEauthorblockA{\textit{National University of Singapore}\\
dcsdjs@nus.edu.sg}
}

\begin{document}

\maketitle
\begingroup
\renewcommand{\thefootnote}{}
\footnotetext[1]{\IEEEauthorrefmark{1} corresponding author}
\endgroup
\begin{abstract}


Recent advances in large language models (LLMs) offer promising potential for automating formal methods. However, applying them to formal verification remains challenging due to the complexity of specification languages, the risk of hallucinated output, and the semantic gap between natural language and formal logic. We introduce PAT-Agent, an end-to-end framework for natural language autoformalization and formal model repair that combines the generative capabilities of LLMs with the rigor of formal verification to automate the construction of verifiable formal models. 
In PAT-Agent, a \emph{Planning LLM} first extracts key modeling elements and generates a detailed plan using semantic prompts, which then guides a \emph{Code Generation LLM} to synthesize syntactically correct and semantically faithful formal models. The resulting code is verified using \emph{the \hl{Process Analysis Toolkit (PAT)} model checker} against user-specified properties, and when discrepancies occur, a \emph{Repair Loop} is triggered to iteratively correct the model using counterexamples. 
To improve flexibility, we built a web-based interface that enables users, particularly non-FM-experts, to describe, customize, and verify system behaviors through user-LLM interactions. Experimental results on 40 systems show that PAT-Agent consistently outperforms baselines, achieving high verification success with superior efficiency.
The ablation studies confirm the importance of both planning and repair components, and the user study demonstrates that our interface is accessible and supports effective formal modeling, even for users with limited formal methods experience.
\end{abstract}


\section{Introduction} \label{sec:intro}

Formal methods provide mathematically rigorous techniques for specifying and verifying the correctness of software and hardware systems. Despite their well-established benefits in safety-critical and concurrent domains, their broader adoption remains limited due to the steep learning curve, lack of accessible tooling, and the complexity of formal languages.

Recent advances in large language models (LLMs), such as OpenAI o3~\cite{openai2025o3} 
and LLaMA~\cite{grattafiori2024llama}, have demonstrated strong capabilities in natural language understanding and code generation. These models present a promising opportunity to bridge the gap between informal requirements and formal specifications, particularly for users without formal methods expertise. By leveraging LLMs, it becomes feasible to lower the barrier to entry for formal modeling and make correctness assurance more broadly accessible.

\noindent \textbf{Challenges.}  
However, applying LLMs to formal model synthesis introduces significant challenges. LLMs are prone to hallucinations and logical inconsistencies~\cite{ji2023survey,yao2023llm}, especially when generating complex artifacts such as process algebra models or temporal logic specifications. Their outputs may violate syntactic rules, diverge from user intent, or overlook critical correctness properties. Existing works on synthesizing formal specifications from natural language instructions have a focus on fine-grained formulae, such as temporal logic~\cite{cosler2023nl2spec, fuggitti2023nl2ltl}. 
There are emerging attempts to synthesize system-level models using LLMs, but these works~\cite{capozucca2025ai, hong2025effectiveness} focus on simpler systems and exhibit limited success, relying solely on direct generation from natural language. In contrast, PAT-Agent integrates structured planning and verification-guided repair, enabling the synthesis of more complex and verifiable models. Meanwhile, traditional rule-based synthesis pipelines are labor-intensive and difficult to generalize across diverse systems and domains. These limitations underscore the need for a unified framework that combines the flexibility and automation capacity of LLMs with the rigor of formal verification to produce reliable, verifiable models at scale.

\noindent \textbf{Our Solution.} 
We propose \textbf{PAT-Agent}, a novel framework for natural language autoformalization and formal code repair. PAT-Agent combines the generative power of LLMs with the rigor of formal verification to automate the construction of verifiable system models. Built on top of the Process Analysis Toolkit (PAT), our approach differs from direct natural language–to–code generation and instead adopts a structured intermediate representation based on semantic prompts. These prompts capture reusable modeling constructs, such as variables and guarded actions, and are parameterized to guide code generation in a controlled, interpretable manner.

The pipeline operates in modular stages. A Planning LLM first analyzes the natural language description and extracts the modeling elements. These elements are further used to populate the structured semantic prompt that guides the generation of a detailed plan for formal code synthesis. A Code Generation LLM then converts the plan into syntactically valid, semantically aligned PAT models. The resulting model is verified using the PAT model checker, and if any user-defined requirements are not satisfied, a Repair Loop is triggered to iteratively revise the model based on counterexamples and failure traces. \hl{This decomposition of tasks across multiple LLMs is motivated by prior works~\cite{jiang2024self,erdogan2025plan}, which show that breaking down complex generation into several sub-tasks can improve effectiveness.}

To support usability, accessibility, and iterative development, we further develop a web-based interface that supports guided modeling, verification, and repair. Users can input natural language descriptions, inspect and edit the extracted intermediate representations, review verification results, and apply automatic corrections through a seamless feedback loop.
All code, datasets, and the interface are 
publically available\footnote{\url{https://github.com/ZuoXinyue/PAT-Agent}}.

This paper makes the following contributions:

\begin{itemize}
\item We propose PAT-Agent, a \emph{fully automated end-to-end} framework for natural language autoformalization and formal code repair. 
It uses a semantic prompt-based decomposition strategy to extract structured modeling elements and incorporates a verification-guided repair loop to iteratively correct invalid models based on model checker feedback. 
An ablation study confirms the effectiveness of both components in achieving high verification success.

\item We conduct a comprehensive evaluation of PAT-Agent on 40 diverse modeling tasks curated from three sources, demonstrating its high effectiveness and efficiency in generating correct and verifiable models compared to direct LLM baselines and alternative model combinations.

\item We design a user-centric interface that enables human-in-the-loop 
development,
complementing the automated pipeline. 
It allows non-experts to 
build
and verify formal system models with minimal effort. 
A user study confirms the usability and accessibility of the interface. 
\end{itemize}
\section{Preliminary}
\label{sec:pre}

This section 
gives an overview 
of the Process Analysis Toolkit (PAT) and its modelling language \emph{CSP\#}.  

\subsection{The PAT Framework}
\hlmeta{Model checking is a method for checking whether a finite-state model of a system meets a given specification. We build our autoformalization framework on PAT (Process Analysis Toolkit~\cite{sun2009pat}), an existing extensible model checker that supports state-space exploration and LTL verification for concurrent and real-time systems. Its underlying formalism, \emph{CSP\#}, is} a dialect of Communicating Sequential Processes (CSP) enriched with imperative C\# statements and typed variables.  
PAT has been successfully applied in domains such as industrial protocols, real-time controllers, and cyber-physical systems\hl{~\cite{sun2013modeling,akande2025ltl}}.
Nevertheless, building a correct model remains an expertise-intensive task. 
Our method tackles this bottleneck by prompting large language models (LLMs) to synthesize CSP\# code directly from informal descriptions, and then verifying it against requirements.

\subsection{Syntax and Semantics of CSP\#}
The subset of CSP\# used in our study can be generated by the grammar:
\[
\begin{array}{lcl}
P & ::= & \textbf{Stop} \mid \textbf{Skip} \mid e \,\rightarrow P 
          \mid e\{\textit{prog}\}\,\rightarrow P \mid P \,\Box\, Q \\[2pt]
  & \mid & \textbf{if}(b)\{P\}\ \textbf{else}\{Q\} \mid P;Q \mid P \parallel_{A} Q \mid P \; ||| \; Q\\[2pt]
\end{array}
\]

A \emph{process} is a labelled transition system (LTS)
\[
  \langle S,\ \mathit{Act},\ \rightarrow,\ s_0 \rangle , \text{ where}
\]
\begin{itemize}[leftmargin=*]
  \item $S$ is the set of program states—valuations of a finite set of typed variables;
  \item $\mathit{Act}$ is the alphabet of \emph{observable events}; internal (silent) actions are denoted $\tau$ and excluded from $\mathit{Act}$;
  \item $\rightarrow \subseteq S \times (\mathit{Act}\cup\{\tau\}) \times S$ is the transition relation;
  \item $s_0 \in S$ is the initial state.
\end{itemize}

A step $s \xrightarrow{\alpha} s'$, with $\alpha$ being an event, communication, or $\tau$, represents an \emph{atomic} state change.  

\noindent\textbf{Event primitives: }%
Besides basic events~$e$, 
CSP\# supports imperative C\# statements written as \verb|{prog}|, which execute atomically and may update multiple variables between two adjacent states.

\noindent\textbf{Compositional operators: }%
Complex systems are built from smaller processes via the operators below:
\begin{itemize}[leftmargin=*]
  \item \emph{Sequential composition} \(\,P;Q\) runs \(P\) 
  and
  then runs \(Q\).
  \item \emph{External choice} \(P \,{\Box}\, Q\) offers the union of initial events; the environment selects a branch.
  \item \emph{Parallel composition} \(P \parallel_{A} Q\) forces synchronization on the alphabet \(A\) and interleaves all other events.
  \item \emph{Interleaving} \(P \; ||| \; Q\) executes both operands independently without event synchronization.
\end{itemize}

\subsection{Assertions and Verification in PAT}
\label{sec:assert}
A directive \verb|#assert| embeds a decision problem about the generated LTS.  We rely on three built-in classes:
\begin{description}[leftmargin=*]
  \item[Deadlock-freedom:] \verb|#assert P deadlockfree| \hl{checks whether} every reachable state enables a transition.
  \item[Reachability:] \verb|#assert P reaches cond| checks whether a state satisfying predicate~\verb|cond| is reachable.
  \item[Linear temporal logic (LTL):] \verb|#assert P| $\models$ F \hl{checks whether} all execution traces of $P$ satisfy the LTL formula~\(F\).
\end{description}

Detailed explanations of the syntax and semantics of CSP\# and PAT’s assertion language are available in the literature~\cite{shi2013}.


\hl{In our pipeline, users specify two inputs: properties in natural language and the expected verification outcomes (VALID or INVALID). These inputs together form the \emph{requirements}. The properties are then autoformalized into corresponding PAT assertions. PAT evaluates the assertions, and the verification results are compared against the expected outcomes. A requirement is satisfied if the two match.}

\section{Approach}\label{sec:approach}

\begin{figure}[h]
 \begin{centering}
  \includegraphics[width=0.79\linewidth]{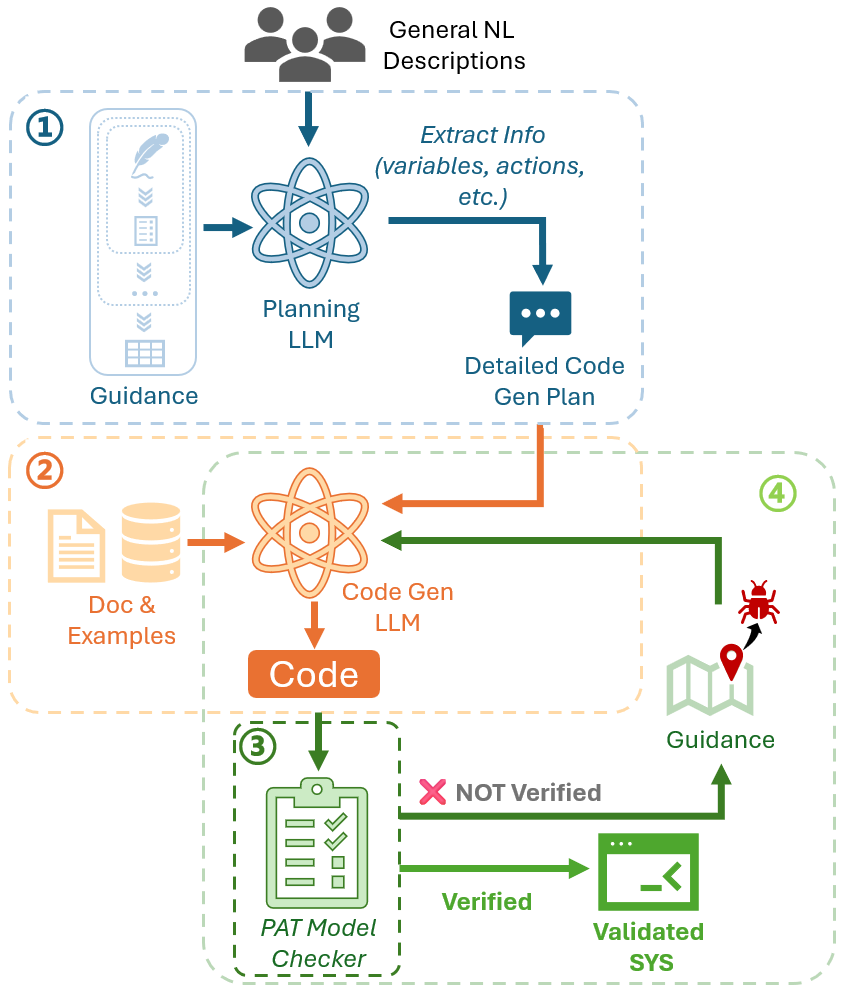}
  \par\end{centering}
 \caption{PAT-Agent Framework Overview.}
 \label{fig:pat_agent}
\end{figure}

We propose \textbf{PAT-Agent}, a novel framework that tightly integrates LLMs with PAT to automate the synthesis and formal verification of system models from natural language specifications. 
As illustrated in Figure~\ref{fig:pat_agent}, the PAT-Agent framework comprises four core components:

\begin{itemize}[leftmargin=*]
    \item \textbf{Planning LLM} analyzes the natural language specification to extract key semantic elements, such as constants, variables, actions, and guarded conditions, and maps them into the parameterized slots of domain-specific semantic prompts, producing a detailed code generation plan that guides formal model construction.

    \item \textbf{Code Generation LLM} converts the detailed plan into formal models written in the PAT specification language, ensuring syntactic correctness and semantic fidelity to the original user intent.

    \item \textbf{Model Checker (PAT)} verifies whether the generated model satisfies the user-specified requirements, including safety, liveness, deadlock-freedom, and produces diagnostic feedback when a requirement is not satisfied.

    \item \textbf{Repair Loop} leverages feedback from PAT, such as counterexamples and violation traces, to iteratively revise the model. While LLMs assist in interpreting the feedback and producing corrections, the pipeline provides targeted guidance to ensure effective and controlled repair.
\end{itemize}

We formalize our pipeline as follows:  
\begin{itemize}[leftmargin=*]
\item $\mathcal{L}_{\mathrm{NL}}$ denotes the set of natural language specifications;
\item $\Pi$ is the space of \emph{planning representations}---structured \hl{semantic prompt templates that are partially instantiated, with placeholders for constants, variables, and actions};
\item $\mathcal{P}$ denotes the detailed model generation plan;
\item $\mathcal{M}$ is the set of well-formed PAT models;
\item \hl{$\mathcal{Q} = \{(\varphi_1,o_1),\ldots,(\varphi_m,o_m)\}$ denotes a finite set of user requirements, where each property $\varphi_i$ is paired with an expected outcome $o_i \in \{\text{VALID}, \text{INVALID}\}$;}
\item $\mathcal{V}$ denotes the verification procedure of PAT;
\item $\mathcal{C}$ \hl{denotes the set of all possible counterexample traces (diagnostic execution sequences) that PAT may return when a requirement is not satisfied;}
\item $\mathcal{R}$ denotes the repair instructions combined with PAT verification results.
\end{itemize}
Given a natural-language specification $N\in\mathcal{L}_{\mathrm{NL}}$ and \hl{a finite
set of requirements $\mathcal{Q}$},
our goal is to construct a \emph{well-formed} PAT model
$M^\star\in\mathcal{M}$ from N such that
\[
    \forall (\varphi,o)\in\mathcal{Q}:\; \mathcal{V}(M^\star,\varphi)=o .
\]

\noindent \textbf{Core Transformations.}
We formalize each pipeline stage as a possibly stochastic transformer:
\[
\begin{aligned}
    \mathcal{T}_{\mathrm{plan}} &:\; \mathcal{L}_{\mathrm{NL}}
        \;\longrightarrow\; \Pi \;\longrightarrow\; \mathcal{P}, \\[2pt]
    \mathcal{T}_{\mathrm{gen}}  &:\; \mathcal{P}
        \;\longrightarrow\; \mathcal{M}, \\[2pt]
    \mathcal{V}                 &:\; \mathcal{M}\times \mathcal{Q}
        \;\longrightarrow\; \{\mathrm{MATCH},\mathrm{MISMATCH}\}^m
            \times\mathcal{C},\\[2pt]
     \mathcal{T}_{\mathrm{repair}} &:\; \mathcal{M}\times\mathcal{C}\times \mathcal{R}
        \;\longrightarrow\; \mathcal{M} .
\end{aligned}
\]

\subsection{Planning LLM}
\label{sec:plan}
To enhance both reliability and interpretability, PAT-Agent creates a semantic prompt-based synthesis strategy in place of end-to-end free-form code generation. 
Rather than directly mapping natural language to formal specifications, the framework introduces an intermediate planning step built around reusable, parameterized modeling prompts. 
Each modeling prompt has four components:

\begin{itemize}[leftmargin=*]
    \item A concise role specification that instructs the LLM to perform a targeted analysis task (e.g., extracting constants and variables from the system description).
    \item Supporting context, including the relevant user-provided natural language description or modeling constructs extracted in earlier stages to inform the current analysis.
    \item Structural guidance for the specific modeling element being extracted. 
    These include:
    \begin{itemize}
        \item Constant: name, value, and description
        \item Variable: name, type, possible values, initial value, and description
        \item Process: name, actions, guard conditions, and associated state changes
    \end{itemize}
    \item Targeted instructions and validation rules for each construct. For example, when processing constants and variables:
    \begin{itemize}
        \item Constants should include three properties: name, integer value, and description. Variables should include five: name, type (e.g., integer, array), possible values (which must be defined as constants if needed), initial value, and description.
        \item All constants must be declared before variables. Names across constants, variables (including their possible values), and processes must be unique and non-conflicting.
    \end{itemize}
    The complete set of instructions is provided
    online~\cite{github}.
\end{itemize}
Formally,
$\mathcal{T}_{\mathrm{plan}}$ leverages \emph{semantic prompts}
$\mathsf{Prompt}_{\mathrm{var}}$,
$\mathsf{Prompt}_{\mathrm{act}}$, \dots to extract
\textsc{(i)}~typed variables and constants, and
\textsc{(ii)}~guarded actions.
The output is JSON-serializable.
These extracted elements are used as inputs to subsequent semantic prompts \hl{($\Pi$)} to generate a detailed natural language annotation --- referred to as the model generation plan $\mathcal{P}$. 
The prompt to generate the coding plan has three components:
(1) role assignment, (2) the previously extracted structured elements, and (3) an instruction to formulate detailed, plan-level annotations that guide formal code generation. 

\subsection{Code-Generation LLM}
Given the detailed modeling plan produced by the Planning LLM that specifies constants, variables, and state transitions, the Code-Generation LLM synthesizes a complete and executable model in the PAT formal specification language. 
\hlmeta{Users also provide properties and expected outcomes that describe the intended system requirements. As system descriptions and requirements are specified as separate inputs in our pipeline, PAT-Agent first constructs the model from the description and then verifies it against the requirements to ensure consistency with the intended system behavior.}

The modeling plan takes the form of a line-by-line natural language annotation that maps directly to formal constructs. To enhance robustness and code quality, we incorporate two forms of prompt augmentation. 
First, we embed syntactic guidance in the form of a compact documentation excerpt, summarizing PAT syntax and common errors observed during early experimentation (e.g., missing semicolons, incorrect process synchronization, malformed assertions). 
These cues help the model avoid structural mistakes and adhere to PAT’s specification conventions. 
Second, we provide semantic guidance via a retrieval-augmented generation (RAG) mechanism. 
For each modeling task, we compare the newly generated plan against a small curated database of \textless{}plan, code\textgreater{} pairs and retrieve the most similar example based on plan content. 
The retrieved example’s plan-code correspondence is then embedded into the prompt, illustrating how a comparable plan was translated into formal PAT code. \hl{If no closely similar plan exists, the retrieved example will serve as a one-shot illustration to guide code structure, syntax, and generation behavior.}

Formally,
$\mathcal{T}_{\mathrm{gen}}$ converts each annotation in
$\mathcal{P}$ into PAT syntax.
Two auxiliary signals improve robustness:
(a)~a \emph{syntax cue} $\sigma_{\mathrm{PAT}}$ summarising common mistakes,
(b)~a retrieval-augmented exemplar $(\mathcal{P}',M')$
with highest cosine similarity to $\mathcal{P}$.


\subsection{Model Checker (PAT)}
Once a formal model is generated, it is automatically verified against user-specified requirements using the PAT model checker. 
For each $\varphi_i$ we invoke PAT’s built-in engines:
deadlock detection, reachability, or LTL.
\hl{PAT returns a raw verification outcome for each $\varphi_i$, which is compared against the user-provided expected outcome $o_i$ to produce a verdict $s_i\!\in\!\{\mathrm{MATCH},\mathrm{MISMATCH}\}$.
Collecting over all $m$ properties yields} the verdict vector $\vec{s}\in\{\mathrm{MATCH},\mathrm{MISMATCH}\}^m$. 
\hl{When $s_i=\mathrm{MISMATCH}$, we also retrieve from PAT} a minimal counter-example trace $C=(t_0\!\to t_1\!\to\cdots\!\to t_\ell)\in\mathcal{C}$, where $t_i, 0\leq i \leq \ell$, are events in the model, \hl{to guide repair.}

\subsection{Verification-Repair Loop}
Starting from an initial model $\mathcal{M}_0=\mathcal{T}_{\mathrm{gen}}(\mathcal{P})$, we perform iterative repair based on verification results:
\[
\boxed{
\begin{aligned}
    (\vec{s}_k,\,C_k)
             &\gets \mathcal{V}(M_k,\mathcal{Q}),\\[2pt]
    \mathcal{M}_{k+1}&\gets
        \begin{cases}
            \mathcal{M}_k, & \text{if } \vec{s}_k\equiv\mathrm{MATCH}^m\\[4pt]
            \mathcal{T}_{\mathrm{repair}}(\mathcal{M}_k,C_k,R_k), & \text{otherwise}.
        \end{cases}
\end{aligned}}
\]
The loop halts at the \hl{first iteration} $k^\star$ \hl{where the verdict vector} $\vec{s}_{k^\star}$ \hl{indicates that all verification results match the expected outcomes specified by the users} (i.e., $\vec{s}_{k^\star}\equiv\mathrm{MATCH}^m$), thereby yielding the verified model $M^\star=M_{k^\star}$. We empirically cap $k$ at $K_{\max}=5$; if no satisfactory model is found within this limit, the user is given the latest model and counterexample trace. \hl{This choice balances repair effectiveness with computational cost, as increasing $K_{\max}$ further yields diminishing returns in our experiments.}

Each iteration leverages structured feedback from PAT in the form of a counterexample trace $C_k$. It is interpreted to generate repair directives $R_k \in \mathcal{R}$, which \textsc{(i)} rank the actions appearing in $C_k$ using a locality-of-fault heuristic, prioritizing nodes closer to the violation point, and \textsc{(ii)} generate targeted edit commands such as \textit{tighten guard of \textsc{start\_driving}}.

Directives are fed back to $\mathcal{T}_{\mathrm{gen}}$ \emph{in-context}, guiding code-level edits without rerunning the full planning pipeline and ensuring locality of change.

\subsection{Interactive Interface}
\label{sec:interface}
While PAT-Agent is fully capable of end-to-end automation, we also provide a user-friendly web-based interface to support controllable and interactive formalization when desired. 
This interface is designed to make formal modeling accessible to non-experts by supporting the entire pipeline through guided, modular interactions. 
It allows users to review and customize system components without requiring prior knowledge of formal specification languages or model checking. Key components of the interface include:

\noindent\textbf{Chatbot:} \hl{Users begin with a high-level natural language description of the system they wish to build, which the chatbot uses to detect reuse opportunities.} It checks the database \hl{of verified systems---stored with both natural language descriptions and formal implementations---and either redirects users to \emph{Verified Example Customization} when a similar system exists, or to \emph{Information Gathering \& Processing} for new synthesis when none match. If the high-level description is ambiguous (i.e., insufficient to determine whether a similar system exists)}, the chatbot requests clarification. \hl{Detailed descriptions and requirements are only collected systematically later in the \emph{Information Gathering \& Processing} stage.}


\noindent\textbf{Verified Example Customization:} 
The interface offers access to historically synthesized and verified systems. \hl{When similarities exist,} users can adjust key parameters \hl{or reuse components}, enabling rapid \hl{adaptation --- similar to code reuse in programming. If no similarities exist, the user proceeds with a fresh synthesis.} To support long-term scalability, PAT-Agent maintains this self-evolving database of verified formal models, indexed by semantic descriptors such as system goals and requirements. Newly synthesized models are automatically added upon successful verification, enriching a growing repository of reusable implementations.

\noindent\textbf{Information Gathering \& Processing:} If a new model is needed, the interface first collects natural language system descriptions from users. It then displays the Planning LLM’s interpreted results in structured tables, each corresponding to a distinct modeling construct. These include constants and variables (with types, value ranges, and initial values), and actions (with associated conditions and state transitions). Users can review and revise the extracted content.

\noindent\textbf{Requirements Specification:} Users specify requirements \hl{(including the intended outcomes)} via a guided input interface with drop-downs and selection tools for defining states using familiar concepts like variable-value pairs. Linear temporal logic formulae can also be created without requiring users to \hl{manually} write formal syntax, \hl{as the system translation internally ensures syntactic correctness. Users can revise their specifications at any time if they notice a misalignment with their intentions.}

\noindent\textbf{\hlmeta{Plan,} Code, and Verification Viewers:} \hlmeta{The Plan and Code viewers allow users to inspect the natural language modeling plan and property descriptions, and the synthesized PAT model and formalized assertions, correspondingly. This enables users to confirm correctness or make edits.} The Verification viewer then presents each requirement with its verification outcome, and, if not satisfied, counterexamples that highlight discrepancies between user expectations and actual system behavior.

\noindent\textbf{Repair Dashboard:} For models that fail to satisfy all specified requirements, the interface displays the current implementation and unsatisfied requirements. 
Users can edit code or trigger automatic repair, and automated updates include human‑readable summaries. \hlmeta{If verification passes, the implementation is guaranteed to meet all user‑confirmed requirements.}

A full demonstration of the interface, along with an illustrative video, is available in our GitHub repository~\cite{github}.
To evaluate the effectiveness and usability of the interface, we conducted a user study involving participants from both formal-methods and non-expert backgrounds, as detailed in Section~\ref{sec:userstudy}. 



\subsection{Running Example}
We illustrate each step with a four-process car model (driver, key, door, motor).
The four processes model four interacting elements --- driver position, key position, door status, and engine behavior --- that collectively define the car's operational logic. 
A valid system design must satisfy key safety and functionality requirements, such as preventing the car from entering a driving state without a licensed driver, avoiding key lock-in scenarios, and ensuring overall deadlock-freedom.

\begin{tcolorbox}[nllisting,title=Description of Engine Behavior]
This process models the motor in the car. The engine can be turned on and off, the car can start and stop driving, the fuel is gradually consumed after driving, and it can be refueled.
\end{tcolorbox}

\begin{figure}[htbp]
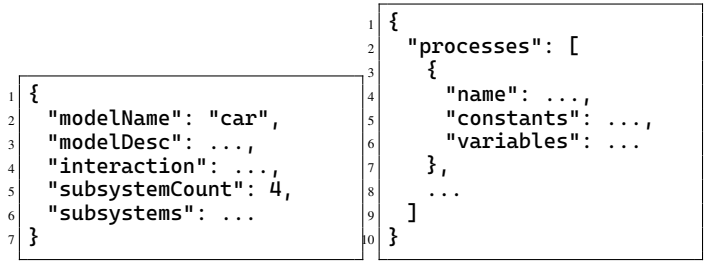

  \centering
  \begin{subfigure}[b]{0.49\columnwidth}
    \begin{lstlisting}[language=json,basicstyle=\ttfamily\footnotesize]
{
  "modelName": "car",
  "modelDesc": ...,
  "interaction": ...,
  "subsystemCount": 4,
  "subsystems": ...
}
    \end{lstlisting}
    \caption{Car System Context 1}
    \label{lst:car-json}
  \end{subfigure}
  \hfill
  \begin{subfigure}[b]{0.46\columnwidth}
    \begin{lstlisting}[language=json,basicstyle=\ttfamily\footnotesize]
{
  "processes": [
    {
      "name": ...,
      "constants": ...,
      "variables": ...
    },
    ...
  ]
}
    \end{lstlisting}
    \caption{Car System Context 2}
    \label{lst:var-json}
  \end{subfigure}

  \caption{Two Context JSONs of the car system in the running example. \hlmeta{These contexts serve as the ``Guidance'' input in Figure~\ref{fig:pat_agent}.}}
  \label{fig:car-jsons}
\end{figure}

\noindent \textbf{Planning with Semantic Prompts.} \hl{Users only need to provide high-level natural language descriptions of the system, such as the engine behavior description shown above. These descriptions are automatically formatted into a} structured supporting context (as illustrated in Figure~\ref{lst:car-json}), which is then used to form the initial semantic prompt for extracting constants and variables.

With the full semantic prompt containing the supporting context, the Planning LLM identifies and extracts relevant constants and variables (e.g., \texttt{\footnotesize carMotion}, \texttt{\footnotesize fuel}, and \texttt{\footnotesize engineStatus}), as demonstrated in Figure~\ref{lst:var-json}.

The extracted elements, along with the relevant system descriptions, are subsequently used as the supporting context in the semantic prompt to reason about possible state transitions, which we refer to as actions. This involves not only identifying explicitly stated behaviors but also applying commonsense reasoning to infer plausible conditions and effects. For instance, even if not stated verbatim, the LLM may infer that the action \textit{start\_driving} should depend on conditions such as the engine being on, sufficient fuel being available, the car being stationary, and the owner being present. These inferred behaviors are essential for generating context-aware and semantically valid formal representations.

\hlmeta{The structured templates within the semantic prompts, such as Figure~\ref{lst:car-json} and Figure~\ref{lst:var-json}, correspond to the input ``Guidance'' in Figure~\ref{fig:pat_agent}, which supports the Planning LLM to analyze systems.} With all modeling constructs extracted, a detailed code generation plan will be formed through the prompt introduced in Section~\ref{sec:plan}.


\noindent \textbf{Plan-Guided Formal Model Generation.}
For example, an annotation in the plan and corresponding PAT code for an action such as \texttt{\footnotesize start\_driving} are as follows:

\begin{cspsharp}{}{}
//If "engineStatus" is "engineOn", "carMotion" is "stop", "fuel" is above 0, and "ownerPos" is "inCar", the action "start_driving" make "carMotion" become "drive" (drive started)
[engineStatus == engineOn && carMotion == stop && (fuel > 0) && ownerPos == inCar] start_driving{carMotion = drive;} -> motor()
\end{cspsharp}

\hlmeta{The ``Doc \& Examples'' in Figure~\ref{fig:pat_agent} consists of syntax documentation and curated \textless{}plan, code\textgreater{} pairs, which guide the Code Generation LLM to produce syntactically correct and semantically consistent models. The full resources are available in our GitHub repository\footnote{\url{https://github.com/ZuoXinyue/PAT-Agent/blob/master/Interface/syntax-dataset.json}, \url{https://github.com/ZuoXinyue/PAT-Agent/blob/master/Interface/database-rag-claude.json}}}.

\noindent \textbf{Verification Feedback and Iterative Repair.} In the car system, key safety and functionality requirements, such as ensuring deadlock-freedom, preventing key lock-in scenarios, and disallowing driving without a licensed driver, are formalized as assertions \hlmeta{and used to test model implementations against user-specified outcomes. Table~\ref{tab:props} shows these assertions alongside both the expected outcomes and the verification results. The implementation meets $P_{1}$ and $P_{2}$, but fails to meet $P_{3}$}.


\begin{table*}[t]
\centering
\caption{\hlmeta{Example Assertions for the Simplified Car System. Each assertion is autoformalized in PAT syntax and paired with the user-specified expected outcome and the verification outcome returned by PAT. 
Green indicates a match between the two. 
Red indicates a mismatch, meaning the requirement is not satisfied.}}
\label{tab:props}
\small
\setlength{\tabcolsep}{6pt}
\renewcommand{\arraystretch}{1.15}
\begin{tabular}{@{}clp{0.61\textwidth}cc@{}}
\toprule
ID & Type & Formalized Assertion & \makecell{User-Specified\\ Outcome} & \makecell{Verification\\ Outcome} \\
\midrule
\textsc{P\textsubscript{1}} & Deadlock &
\texttt{\footnotesize \#assert car deadlockfree;} &
{VALID} &
{\textcolor{darkgreen}{VALID}} \\
\midrule
\textsc{P\textsubscript{2}} & Reachability &
\makecell[l]{\texttt{\footnotesize \#define keyInsideDoorLocked (key == inCar \&\& door == locked}\\
\texttt{\footnotesize \ \ \ \ \ \ \ \ \ \ \ \ \ \ \ \ \ \ \ \ \ \ \ \ \ \ \ \ \ \ \ \ \ \ \ \ \ \ \&\& ownerPos != inCar);}\\
\texttt{\footnotesize \#assert car reaches keyInsideDoorLocked;}} &
{INVALID} &
{\textcolor{darkgreen}{INVALID}} \\
\midrule
\textsc{P\textsubscript{3}} & Reachability &
\makecell[l]{\texttt{\footnotesize \#define driveWithoutDriver (carMotion==drive \&\& ownerPos != inCar);}\\
\texttt{\footnotesize \#assert car reaches driveWithoutDriver;}} &
{INVALID} &
{\textcolor{darkred}{VALID}} \\
\bottomrule
\end{tabular}
\end{table*}

\hlmeta{This requirement is not satisfied} due to potential ambiguities in natural language descriptions or minor generation inaccuracies (e.g., missing preconditions or incorrect behavior specifications), which can cause the initial model to fail some verification criteria. Not satisfying $P_{3}$ allows the car to enter a driving state without a person detected in the cockpit, a clear safety and regulatory violation. When such failures occur, the PAT model checker returns detailed diagnostic feedback in the form of violation traces, which act as counterexamples.
These traces identify specific execution paths or conditions that trigger the violation, helping to isolate the root cause. For example, a trace revealing why \textsf{\footnotesize driveWithoutDriver} was marked \texttt{\footnotesize VALID} for the current system implementation, rather than the desired \texttt{\footnotesize INVALID}, is as below:

\begin{cspsharp}{}{}
<init -> owner_approach -> pickup_key_by_owner -> 
unlock_door_by_owner -> open_door -> owner0_enter -> 
start_engine_by_owner0 -> owner0_exit -> start_driving>
\end{cspsharp}

To revise the model to satisfy the safety requirement, PAT-Agent transforms the diagnostic feedback from PAT into a structured repair strategy and executes it through an iterative repair loop. Instead of re-running the full planning stage, it constructs an augmented prompt that \hl{combines information extracted from the violation trace (e.g., the actions or transitions leading to failure) with repair heuristics tailored to the property type of the unsatisfied requirement (e.g., tightening guard conditions for safety or loosening them for liveness), thereby giving targeted} revision guidance to the Code Generation LLM.

When a trace implicates multiple actions, the prompt prioritizes those more likely responsible \hl{using two heuristics: actions appearing later in the violation trace (i.e., closer to the failure point) and actions occurring more frequently.} For example, in the above trace, \texttt{\footnotesize start\_driving} \hl{appears last and is therefore considered more likely responsible}, while other implicated actions are still considered with lower likelihood.

If \texttt{\footnotesize start\_driving} is wrongly triggered when no driver is present, the prompt suggests tightening its preconditions (e.g., adding a check for driver position). If deadlock issues occur, it may recommend ensuring every state has at least one outgoing transition or relaxing overly strict conditions on related actions. This guided, localized repair preserves correct model components while accelerating convergence toward a verifiable implementation.


\section{Evaluation}

We conduct a comprehensive evaluation of \textbf{PAT-Agent} to assess its effectiveness and efficiency in synthesizing verified models from natural language descriptions. 
Our experiments are designed to answer the following research questions:

\begin{itemize}[leftmargin=*]
    \item \textbf{RQ1:} How effective is the proposed PAT-Agent pipeline in generating formally verified systems from natural language descriptions, compared to alternative generation strategies?
    \item \textbf{RQ2:} What is the impact of the key components comprising the Planning LLM and the Repair Loop on the overall performance of the PAT-Agent pipeline?
    \item \textbf{RQ3:} How does the processing time vary across different stages of the PAT-Agent pipeline and different model configurations?
\end{itemize}

\subsection{Datasets and Baselines}

We evaluate PAT-Agent on three datasets: \textbf{PAT} (26 pairs of natural language descriptions and requirements collected from the PAT library), 
\textbf{A4F} (8 pairs adapted from the Alloy4Fun dataset~\cite{macedo2019sharing}), and \textbf{UCS} (6 pairs curated from the CSP textbook Understanding Concurrent Systems~\cite{roscoe2010understanding}), totaling 40 modeling examples.
We curated the examples that can be formalized in the syntax defined in \autoref{sec:pre}.
Table~\ref{tab:datasets} summarizes the datasets used in our evaluation. 
It reports the number of system instances and total number of verification assertions, as well as two statistics computed over the final verified models: the average of each system’s maximum number of visited states and the average of each system’s maximum number of transitions observed during verification. 

\begin{table}[t!]
\centering
\caption{Dataset Statistics.}
\label{tab:datasets}
\begin{tabular}{@{}ccccc@{}}
\toprule
\textbf{Dataset} & \textbf{\#Instances} & \textbf{\#Assertions} & \textbf{States} & \textbf{Transitions} \\ \midrule
PAT              & 26         & 74         & 101253.62   & 342084.42                   \\
UCS              & 6          & 19         & 1164.17   & 14472.83                   \\
A4F              & 8          & 40         & 1223.00   & 7008.50                \\
Overall          & 40         & 133        & 66234.07   & 225927.50              \\ \bottomrule
\end{tabular}
\end{table}



In our study, we adopt OpenAI’s \textit{o3-mini-2025-01-31}\footnote{\url{https://platform.openai.com/docs/models/o3-mini}} as the Planning LLM (referred to as o3) and Anthropic’s \textit{claude-3-7-sonnet-20250219}\footnote{\url{https://docs.anthropic.com/en/docs/about-claude/models/overview}} as the Code Generation LLM (referred to as Claude). \hl{We denote this configuration as \textless{}o3, Claude\textgreater{}, which serves as} the default setup of our PAT-Agent pipeline. 
To benchmark the effectiveness and efficiency of alternative model pairings, we also experiment with \textit{DeepSeek-R1}\footnote{\url{https://api-docs.deepseek.com/news/news250120}} (referred to as R1).

\subsection{Metrics}

To evaluate the effectiveness of our framework, we employ three metrics commonly used in software engineering research~\cite{wang2022compilable, chen2021evaluating, hendrycks2021measuring}: Compilation Success Rate (CSR), Full-Pass Rate (FPR), and Average Pass Rate (APR). \hlmeta{In our setting, ``pass'' means that the verification result produced by PAT matches the expected outcome specified by the user for an assertion, and ``full-pass'' means that a system satisfies all of its specified requirements. The metrics are} defined as follows:

\[
\begin{aligned}
\mathrm{CSR}
  &= \frac{\lvert S_{\mathrm{succ}}\rvert}{N},
&
\mathrm{FPR}
  &= \frac{\lvert S_{\mathrm{full}}\rvert}{N},
&
\mathrm{APR}
  &= \frac{\sum_{i=1}^{N} p_{i}}{\sum_{i=1}^{N} A_{i}}
\end{aligned}
\]

Let \(N\) denote the total number of systems for which formal models are generated.  \(\mathrm{CSR}\) measures the proportion of these systems that compile without syntax errors, where \(S_{\mathrm{succ}}\subseteq N\) denotes those successfully compiled.

\(\mathrm{FPR}\) captures the fraction of systems that pass all their assertions, with \(S_{\mathrm{full}}\subseteq N\) denoting such systems.

\(\mathrm{APR}\) offers a fine-grained view of assertion-level correctness. Let \(p_i\) be the number of assertions passed by system \(i\), and \(A_i\) the total assertions it must satisfy; APR then reflects the overall proportion of satisfied assertions across all systems --- even when some systems fail a subset of checks.





\subsection{Results of RQ1: Effectiveness of Formal Model Generation}

\noindent \textbf{Settings.}
To evaluate the effectiveness of our pipeline in generating formally verified models from natural language descriptions, we compare its performance against direct LLM-based generation methods, as well as against five alternative model combinations using the same pipeline structure. 
The baseline methods include individual LLMs (o3, Claude, R1) used to generate code directly without a structured planning step or repair loop. 
These direct generations are provided with the same RAG-extracted exemplars and syntax documentation as our pipeline to ensure a fair comparison.

For model combination comparisons, we explore five additional pairings for the planning and code generation stages, resulting in the variants shown in Table~\ref{tab:RQ1}. 
Our chosen configuration --- o3 for planning and Claude for code generation --- serves as the default setting for PAT-Agent.

\begin{table}[t!]
\centering
\caption{Performance against Direct Generation with LLMs.}
\label{tab:directGen}
\begin{tabular}{@{}llccc@{}}
\toprule
\textbf{Dataset} & \multicolumn{1}{l|}{\textbf{Model}} & \textbf{CSR}  & \textbf{FPR} & \textbf{APR}   \\ \midrule
\multirow{4}{*}{PAT}              & \multicolumn{1}{l|}{R1}                        & 0.3846                & 0.3462             & 0.3243             \\
                                  & \multicolumn{1}{l|}{o3}                              & 0.7308                & 0.5769             & 0.7162             \\
                                  & \multicolumn{1}{l|}{Claude}                          & 0.7308                & 0.5385             & 0.6892             \\
                                  & \multicolumn{1}{l|}{\textbf{Ours}}              & \textbf{1.0000}       & \textbf{1.0000}    & \textbf{1.0000}    \\ \midrule
\multirow{4}{*}{UCS}              & \multicolumn{1}{l|}{R1}                        & 0.3333                & 0.3333             & 0.2632             \\
                                  & \multicolumn{1}{l|}{o3}                              & 0.6667                & 0.5000             & 0.5789             \\
                                  & \multicolumn{1}{l|}{Claude}                          & 0.6667                & 0.1667             & 0.4211             \\
                                  & \multicolumn{1}{l|}{\textbf{Ours}}              & \textbf{1.0000}       & \textbf{1.0000}    & \textbf{1.0000}    \\ \midrule
\multirow{4}{*}{A4F}            & \multicolumn{1}{l|}{R1}                        & 0.6250                & 0.5000             & 0.5750             \\
                                  & \multicolumn{1}{l|}{o3}                              & 0.7500                & 0.3750             & 0.5250             \\
                                  & \multicolumn{1}{l|}{Claude}                          & 0.8750                & 0.5000             & 0.6250             \\
                                  & \multicolumn{1}{l|}{\textbf{Ours}}              & \textbf{1.0000}       & \textbf{1.0000}    & \textbf{1.0000}    \\ \midrule
\multirow{4}{*}{Overall}          & \multicolumn{1}{l|}{R1}                        & 0.4250                & 0.3750             & 0.3910             \\
                                  & \multicolumn{1}{l|}{o3}                              & 0.7250                & 0.5250             & 0.6391             \\
                                  & \multicolumn{1}{l|}{Claude}                          & 0.7500                & 0.4750             & 0.6316             \\
                                  & \textbf{Ours}                                   & \textbf{1.0000}       & \textbf{1.0000}    & \textbf{1.0000}    \\ \bottomrule
\end{tabular}
\end{table}

\begin{table}[t!]
\centering
\caption{Verification Results of Formal Models Generated by Different Pipeline Configurations. Each entry in the Pipeline column is denoted as \textless{}model\textsubscript{plan}, model\textsubscript{gen}\textgreater{}, where model\textsubscript{plan} is used for planning and model\textsubscript{gen} for code generation. Best results are bolded; second-best results are underscored.}
\label{tab:RQ1}
\begin{tabular}{@{}ll|ccc@{}}
\toprule
\textbf{Dataset} & \textbf{Pipeline}                                        & \textbf{CSR}  & \textbf{FPR} & \textbf{APR}   \\ \midrule
\multirow{6}{*}{PAT}     & \textbf{\textless{}o3, Claude\textgreater{}} & \textbf{1.0000}       & \textbf{1.0000}    & \textbf{1.0000}    \\
                         & \underline{\textless{}R1, Claude\textgreater{}}          & \underline{0.8846}                & \underline{0.8462}             & \underline{0.8514}             \\
                         & \textless{}R1, R1\textgreater{}              & 0.6923                & 0.6538             & 0.6081             \\
                         & \textless{}R1, o3\textgreater{}              & 0.6923                & 0.6923             & 0.6351             \\
                         & \textless{}o3, R1\textgreater{}              & 0.8462                & 0.8077             & 0.7297             \\
                         & \textless{}o3, o3\textgreater{}              & 0.8846                & 0.8077             & 0.8649             \\ \midrule
\multirow{6}{*}{UCS}     & \textbf{\textless{}o3, Claude\textgreater{}} & \textbf{1.0000}       & \textbf{1.0000}    & \textbf{1.0000}    \\
                         & \textbf{\textless{}R1, Claude\textgreater{}} & \textbf{1.0000}       & \textbf{1.0000}    & \textbf{1.0000}    \\
                         & \textless{}R1, R1\textgreater{}              & 0.6667                & 0.6667             & 0.6842             \\
                         & \textless{}R1, o3\textgreater{}              & 0.6667                & 0.5000             & 0.6316             \\
                         & \textless{}o3, R1\textgreater{}              & 1.0000                & 0.8333             & 0.9474             \\
                         & \textless{}o3, o3\textgreater{}              & 0.6667                & 0.6667             & 0.6842             \\ \midrule
\multirow{6}{*}{A4F}   & \textbf{\textless{}o3, Claude\textgreater{}} & \textbf{1.0000}       & \textbf{1.0000}    & \textbf{1.0000}    \\
                         & \textless{}R1, Claude\textgreater{}          & 0.8750                & 0.7500             & 0.8000             \\
                         & \textless{}R1, R1\textgreater{}              & 0.8750                & 0.7500             & 0.8250             \\
                         & \textless{}R1, o3\textgreater{}              & 0.8750                & 0.7500             & 0.8000             \\
                         & \textless{}o3, R1\textgreater{}              & 0.8750                & 0.7500             & 0.8250             \\
                         & \underline{\textless{}o3, o3\textgreater{}}              & \underline{1.0000}                & \underline{0.8750}             & \underline{0.9250}             \\ \midrule
\multirow{6}{*}{Overall} & \textbf{\textless{}o3, Claude\textgreater{}} & \textbf{1.0000}       & \textbf{1.0000}    & \textbf{1.0000}    \\
                         & \underline{\textless{}R1, Claude\textgreater{}}          & \underline{0.9000}                & \underline{0.8500}             & \underline{0.8571}             \\
                         & \textless{}R1, R1\textgreater{}              & 0.7250                & 0.6750             & 0.6842             \\
                         & \textless{}R1, o3\textgreater{}              & 0.7250                & 0.6750             & 0.6842             \\
                         & \textless{}o3, R1\textgreater{}              & 0.8750                & 0.8000             & 0.7895             \\
                         & \textless{}o3, o3\textgreater{}              & 0.8750                & 0.8000             & 0.8571             \\ \bottomrule
\end{tabular}
\end{table}

\noindent \textbf{Results.}
From Table~\ref{tab:directGen}, we observe that direct LLM-based generations are only able to obtain fully verified implementations for 37.5\% to 52.5\% of the systems, despite structured prompting conditions. Furthermore, the strongest individual model still fails to compile in 25\% of the cases.
These failures are often due to incomplete condition handling, misinterpretation of requirements, or improper modeling abstractions for more complex systems.


In contrast, our pipeline achieves a 100\% full-pass rate across all 40 systems. 
This result does not indicate flawless first-attempt generations, rather, it highlights the effectiveness of the integrated repair loop, which iteratively incorporates counterexamples and verification feedback to correct model errors until all specified requirements are satisfied, as shown in \autoref{tab:RQ2_repair}.
A detailed breakdown of first-attempt and per-round performance is available in our GitHub repository~\cite{github}, showing that multiple repair iterations were often necessary.

Table~\ref{tab:RQ1} further demonstrates that our \textless{}o3, Claude\textgreater{} combination consistently outperforms all alternative model pairings across datasets. 
Across all datasets, the \textless{}o3, Claude\textgreater{} configuration achieves 1.0000 in compilation success rate, full-pass rate, and average pass rate, indicating that the generated systems not only compile successfully but also satisfy all user-specified requirements after repair. 
Other combinations show varying degrees of degradation. 
While \textless{}R1, Claude\textgreater{} performs relatively well overall (i.e., 0.9000 compilation and 0.8500 full-pass rate), it still fails to match the consistency of our pipeline. R1, when used as the planner, sometimes produces outputs that are ambiguous or insufficiently differentiated. 
For instance, it may fail to clearly separate constant identifiers from process names. 
This leads to conflicting definitions that confuse the Code Generation LLM, resulting in verification failures despite Claude’s strong generative ability.


Beyond performance, we also note practical advantages in instruction-following: o3 reliably adheres to structured output formats (e.g., valid JSON), which simplifies integration into our user-facing interface. 
In contrast, R1 frequently ignores prompt constraints or produces malformed outputs, making it less suitable for interactive settings where predictable formatting is crucial for usability.

\begin{figure*}[h]
 \begin{centering}
  \includegraphics[width=0.98\linewidth]{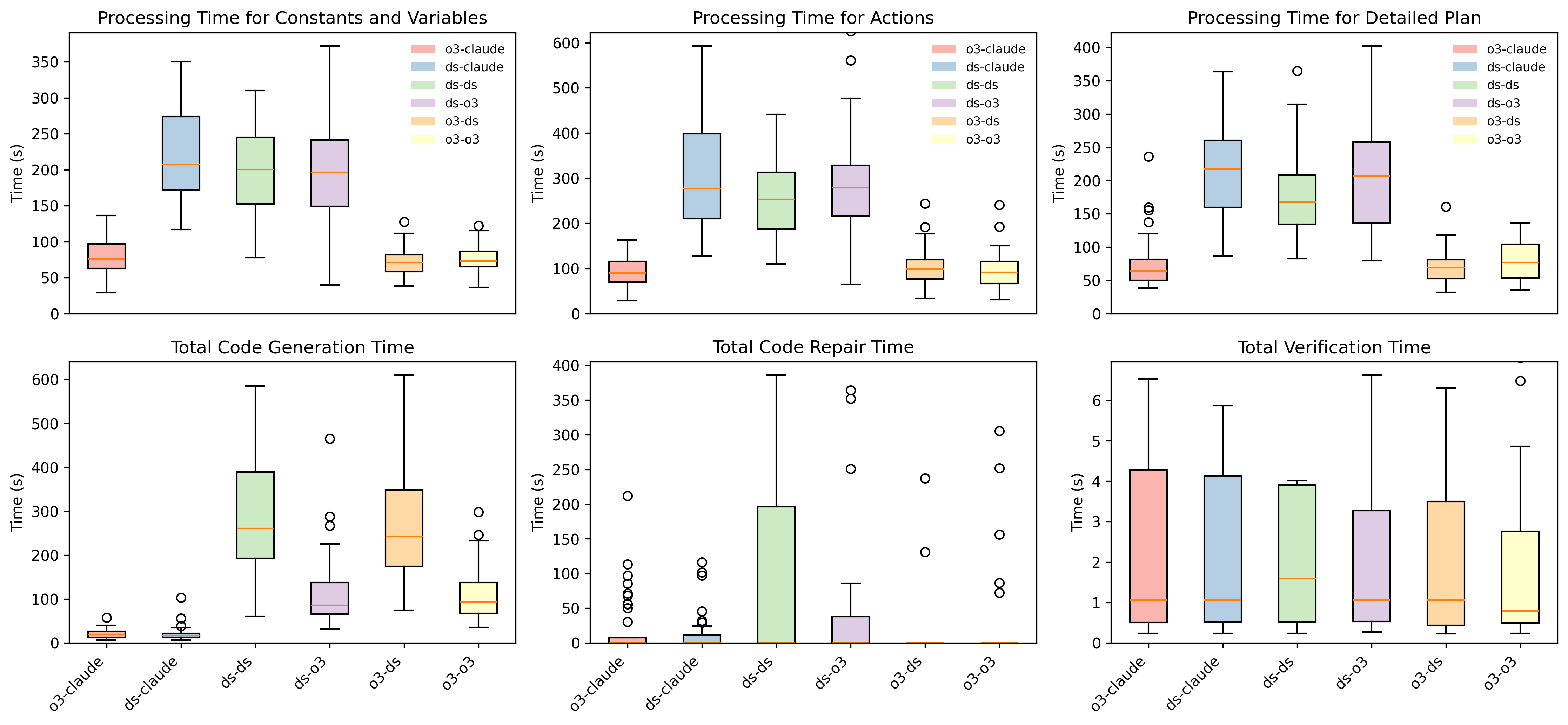}
  \par\end{centering}
 \caption{Time Cost for Each Step across Pipelines.}
 \label{fig:times}
\end{figure*}

\subsection{Results of RQ2: Impact of Components}
\label{sec:ablation}

\noindent \textbf{Settings.}
We perform ablations by disabling two key components of the PAT-Agent pipeline: (a) the Planning LLM module and (b) the Verification-Guided Repair Loop. 
Our goal is to assess how much each component contributes to the final verification performance. 
We evaluate these variants using the same set of metrics as in RQ1 --- compilation success rate, full-pass rate, and average pass rate --- across all datasets.

\begin{table}[t!]
\centering
\caption{Ablation Study.}
\label{tab:RQ2}
\begin{tabular}{@{}ll|ccc@{}}
\toprule
\textbf{Dataset} & \textbf{Method} & \textbf{CSR}  & \textbf{FPR} & \textbf{APR}   \\ \midrule
\multirow{4}{*}{PAT}              & \textbf{Full Pipeline}           & \textbf{1.0000}       & \textbf{1.0000}    & \textbf{1.0000}    \\
                                  & Without Repair Loop              & 1.0000                & 0.8077             & 0.8378             \\
                                  & Without Planning Model               & 0.7308                & 0.5769             & 0.7162             \\
                                  & Without Both Components                & 0.7308                & 0.5385             & 0.6892             \\ \midrule
\multirow{4}{*}{UCS}              & \textbf{Full Pipeline}           & \textbf{1.0000}       & \textbf{1.0000}    & \textbf{1.0000}    \\
                                  & Without Repair Loop              & 1.0000                & 0.8333             & 0.9474             \\
                                  & Without Planning Model               & 0.6667                & 0.5000             & 0.6316             \\
                                  & Without Both Components                & 0.6667                & 0.1667             & 0.4211             \\ \midrule
\multirow{4}{*}{A4F}            & \textbf{Full Pipeline}           & \textbf{1.0000}       & \textbf{1.0000}    & \textbf{1.0000}    \\
                                  & Without Repair Loop              & 1.0000                & 0.5000             & 0.6750             \\
                                  & Without Planning Model               & 0.8750                & 0.7500             & 0.9000             \\
                                  & Without Both Components                & 0.8750                & 0.5000             & 0.6250             \\ \midrule
\multirow{4}{*}{Overall}          & \textbf{Full Pipeline}           & \textbf{1.0000}       & \textbf{1.0000}    & \textbf{1.0000}    \\
                                  & Without Repair Loop              & 1.0000                & 0.7500             & 0.8045             \\
                                  & Without Planning Model               & 0.7500                & 0.6000             & 0.7594             \\
                                  & Without Both Components                & 0.7500                & 0.4750             & 0.6316             \\ \bottomrule
\end{tabular}
\end{table}

\noindent We have the following variants:
\begin{itemize}[leftmargin=*]
\item \textbf{Full Pipeline}: The complete PAT-Agent configuration, with both Planning and Repair components enabled.
\item \textbf{Without Planning Model}: The detailed code generation plan produced by the Planning LLM is removed. Instead, the Code Generation LLM receives only the original natural language description as input. Other components, such as RAG-based example retrieval and syntax documentation, are still included to ensure fair comparison.
\item \textbf{Without Repair Loop}: The verification-guided repair process is disabled. The system only evaluates the initial model generated based on the planning output, without iterative correction based on verification feedback.
\item \textbf{Without Both Components}: Both the Planning LLM and the Repair Loop are removed. The Code Generation LLM directly operates on the raw natural language input in a single-pass manner, but still benefits from RAG examples and syntax documentation.
\end{itemize}

\begin{table}[]
\centering
\caption{Breakdown of Repair Loop Impact.}
\label{tab:RQ2_repair}
\begin{tabular}{@{}ll|ccc@{}}
\toprule
\textbf{Dataset}         & \textbf{Repair Round} & \textbf{CSR} & \textbf{FPR} & \textbf{APR} \\ \midrule
\multirow{4}{*}{PAT}     & Round 0               & 1.0000                            & 0.8077                  & 0.8378                     \\
                         & Round 1               & 1.0000                            & 0.8077                  & 0.8784                     \\
                         & Round 2               & 1.0000                            & 0.8462                  & 0.9324                     \\
                         & \textbf{Round 5}      & \textbf{1.0000}                   & \textbf{1.0000}         & \textbf{1.0000}            \\ \midrule
\multirow{4}{*}{CSP}     & Round 0               & 1.0000                            & 0.8333                  & 0.9474                     \\
                         & Round 1               & 1.0000                            & 0.8333                  & 0.9474                     \\
                         & Round 2               & 1.0000                            & 1.0000                  & 1.0000                     \\
                         & \textbf{Round 5}      & \textbf{1.0000}                   & \textbf{1.0000}         & \textbf{1.0000}            \\ \midrule
\multirow{4}{*}{Alloy}   & Round 0               & 1.0000                            & 0.5000                  & 0.6750                     \\
                         & Round 1               & 1.0000                            & 0.6250                  & 0.8250                     \\
                         & Round 2               & 1.0000                            & 0.8750                  & 0.9750                     \\
                         & \textbf{Round 5}      & \textbf{1.0000}                   & \textbf{1.0000}         & \textbf{1.0000}            \\ \midrule
\multirow{4}{*}{Overall} & Round 0               & 1.0000                            & 0.7500                  & 0.8045                     \\
                         & Round 1               & 1.0000                            & 0.7750                  & 0.8722                     \\
                         & Round 2               & 1.0000                            & 0.8750                  & 0.9549                     \\
                         & \textbf{Round 5}      & \textbf{1.0000}                   & \textbf{1.0000}         & \textbf{1.0000}            \\ \bottomrule
\end{tabular}
\end{table}

\noindent \textbf{Results.} 
Table~\ref{tab:RQ2} shows that removing either the Planning LLM or the Repair Loop causes a significant drop in verification performance, with the most severe degradation occurring when both are disabled.

Removing the Repair Loop alone still maintains 100\% compilation but causes the full-pass rate to drop to 0.7500 and average pass rate to 0.8045, highlighting that many models require verification-guided adjustments to satisfy all user-defined assertions.
To further illustrate the impact of the repair loop, we provide a breakdown of performance across individual repair rounds in Table~\ref{tab:RQ2_repair}. This includes the initial generation (Round 0), intermediate repair rounds (1–2), and the final round (Round 5), demonstrating how iterative feedback progressively improves model correctness.
The results show that the performance increases and finally goes to 100\% during the repair loop.

On the other hand, excluding the Planning LLM has a substantial negative impact on both syntactic and semantic quality. 
The compilation success rate drops to 0.7500, indicating that without structured, semantics-aware planning, the same Code Generation LLM struggles to produce even syntactically valid PAT code in a quarter of the cases. 
Moreover, the full-pass rate drops to 0.6000, and the average pass rate falls to 0.7594, highlighting that even when the code compiles, it might fail to satisfy the requirements after repair. 
These results demonstrate the importance of the Planning LLM in structuring the input for code generation, ensuring not only syntax correctness but also alignment with user intent.


Overall, these results affirm the importance of both the Planning LLM and the Repair Loop: planning provides high-quality structural guidance, while repair ensures final verification success through iterative corrections.

\subsection{Results of RQ3: Time Efficiency}

\noindent \textbf{Settings.}
We evaluate the time efficiency of different pipeline configurations by recording the wall-clock time for each processing step, including constant and variable analysis, action extraction, instruction generation (all part of planning), code generation, repair, and verification. 
These measurements exclude user interaction time, which is assessed separately in the user study (Section~\ref{sec:userstudy}). 
Timing is computed only for systems that are eventually verified, so the number of samples varies across configurations. The detailed statistics are provided in our GitHub repository~\cite{github}. 
Figure~\ref{fig:times} presents box plots of the time taken by each processing step, with y-axes capped to improve visual clarity.

\noindent \textbf{Results.} 
Our pipeline configuration \textless{}o3, Claude\textgreater{} achieves not only the highest verification performance (as shown in RQ1) but also competitive efficiency across all stages. 
o3 consistently outperforms R1 in planning steps, reflecting both faster response time and more effective structure extraction. 
Claude exhibits the lowest latency in code generation among all models, confirming its fluency in producing valid PAT code.

For repair, the \textless{}o3, Claude\textgreater{} pipeline maintains low and stable runtimes, indicating that many systems are verified with only minor corrections. \textless{}R1, R1\textgreater{} and \textless{}R1, o3\textgreater{} exhibit generally longer repair times across their verified systems. 
In contrast, \textless{}o3, R1\textgreater{} and \textless{}o3, o3\textgreater{} show near-zero repair time for most of their successfully verified systems, suggesting that repair was often unnecessary. 
However, this efficiency must be considered alongside their lower overall verification coverage (as shown in RQ1), which means fewer systems benefited from repair in the first place. 
A small number of long-tail cases in these pipelines show elevated repair time, indicating that correction, when required, can occasionally be time-consuming. Verification time remains consistently low across all configurations, confirming that model checking contributes minimally to overall latency.

When aggregating the total time required to generate a verifiable system, \textless{}o3, Claude\textgreater{} achieves the lowest median runtime at 4.34 minutes per system. 
These results confirm that PAT-Agent achieves strong end-to-end performance with high time efficiency.
\section{User Study}
\label{sec:userstudy}

To evaluate the usability and effectiveness of PAT-Agent's interactive mode in practical modeling scenarios, particularly for those without formal methods expertise, we conducted a user study centered around natural language-based system modeling and formal verification tasks. 

\noindent \textbf{Settings.}
We recruited 20 participants with 2–8 years of computer science background. Among them, 30\% had prior experience with PAT or CSP\#, ranging from 1 to 5 years, ensuring the study included both novices and semi-experienced users.
Participants were randomly assigned into two groups --- Control Group (CG) and Experimental Group (EG) --- with 70\% novices in each group, and were numbered P1–P10 (CG) and P11–P20 (EG), respectively.
All participants received a 10-minute tutorial on PAT and CSP\#. 

Both groups were also provided with
(1) Clear natural language descriptions of the target systems and corresponding requirements.
(2) The official PAT documentation.
(3) A list of common modeling errors.
(4) A well-formed PAT model example for reference.


Control Group (CG): 
Participants in the control group were instructed to use any resource (including LLM) except for PAT-Agent 
to generate PAT models from the given natural language system descriptions. They were encouraged to leverage the provided grammar guides and example errors to construct effective prompts and repair the generated code. 

Experimental Group (EG): 
Participants in the experimental group used our PAT-Agent interface. Following the shared tutorial, they were given an additional walkthrough session that demonstrated how to use PAT-Agent.
During the task, EG participants used PAT-Agent exclusively.

Modeling Tasks:
To account for the expected difficulty faced by participants without prior PAT experience, 
we selected one relatively simple system from each of our three datasets, along with one more complex task.
\begin{itemize}
    \item System 1 models a file system, where each file can be created, trashed, and protected.
    \item System 2 represents a trading scenario in which a buyer can purchase an item from a merchant if the offered price meets or exceeds the merchant’s expectation.
    \item System 3 captures the classic missionaries and cannibals river-crossing puzzle.
    \item System 4 is a complex car system alluded in Section~\ref{sec:approach}.
\end{itemize}

\noindent \textbf{Metrics.}
We evaluated the effectiveness and usability of PAT-Agent using both objective performance indicators and user feedback, as explained below.

\noindent{\emph{Task Completion Time}:}
The average 
time taken to complete
modeling,
measured in minutes and averaged across all systems. 

\noindent{\emph{Assertion Accuracy}:}
The percentage of assertions written by each participant that were both syntactically correct and semantically aligned with the intended properties. 

\noindent{\emph{System Accuracy}:}
The proportion of reference (ground-truth) assertions for each task that were correctly verified --- i.e., the participant-generated model produced the expected verification outcome (\texttt{\footnotesize VALID} or \texttt{\footnotesize INVALID}) for each property. 

\noindent{\emph{Statistical Testing}:}
We 
used
the Mann–Whitney U test~\cite{mann1947test} to compare 
CG and EG.
This nonparametric test requires no assumption of normally distributed data and is 
suited to small sample sizes and ordinal measures. All comparisons were two-tailed, with a significance threshold of $\alpha = 0.05$.

\begin{table}[]
\centering

\caption{Comparison of 
CG and EG
Performance. Statistical significance: * ($p<0.05$), ** ($p<0.01$), *** ($p<0.001$).} 
\begin{tabular}{@{}lccc@{}}
\toprule
Metric             & EG Mean & CG Mean & p-value  \\ \midrule
Time (min)         & 12.85$^{*}$   & 17.11   & 1.16E-02 \\
Assertion Accuracy & 0.9958$^{***}$  & 0.7500  & 1.85E-07 \\
System Accuracy    & 0.9688$^{***}$  & 0.6633  & 2.66E-05 \\ \bottomrule
\end{tabular}
\label{tab:userstudy}
\end{table}

\noindent \textbf{Results.}
Full user study outcomes are available in our GitHub repository~\cite{github}. We present the summarized user-study outcomes in Table~\ref{tab:userstudy}, reporting the mean values for each metric in both control and experimental groups over the four system tasks, together with the corresponding significance results. 
Our principal findings are as follows: 

\begin{itemize}[leftmargin=*]
    \item \textbf{Task Efficiency}: 
    EG users, using PAT-Agent, completed tasks faster than CG users, showing improved efficiency and reduced cognitive overhead. RQ3 shows that our pipeline averages 4.34 minutes to generate a verifiable system. The difference suggests users spend time validating intermediate outputs and interpreting verification results via the interface.
    
    \item \textbf{Assertion Accuracy:} EG users achieved near-perfect assertion accuracy (0.9958), significantly outperforming CG. This reflects the framework’s effectiveness in helping users formalize system properties correctly, which is an essential step for ensuring overall model correctness.
    \item \textbf{System Accuracy:} Models generated by EG users achieved significantly higher system accuracy compared to CG. For simpler tasks, EG consistently reached 1.0000 accuracy across all users. In CG, users with a formal methods background tended to perform more consistently across systems, while those without such experience struggled to produce verifiable models, particularly on system 4. In contrast, PAT-Agent enabled consistent success among EG users regardless of prior experience.
    
    \item \textbf{User Feedback:} To evaluate the interface design, we gathered user feedback on its clarity and ease of use. Several CG participants reported challenges in prompting LLMs and writing correct PAT code. In contrast, EG participants found the modular layout and structured tables intuitive, helping them understand how formal models are constructed. All EG users highlighted that the interface made it easy to inspect and modify specific model components. For example, P19 noted that it allowed convenient review of variable declarations and controlled adjustments. These responses suggest the interface effectively balances automation with user control.
\end{itemize}

\section{Discussion}
Auto-formalization systems that translate raw natural language directly into a target formal language have been proposed for mainstream notations such as TLA$^{+}$~\cite{lamport2002specifying}, Z~\cite{spivey1992z}, or Isabelle theories~\cite{nipkow2002isabelle}.
In contrast, CSP\# is comparatively niche and lacks public corpora or prior end-to-end translators on which a strong auto-formalization baseline could be built.  
To the best of our knowledge, no existing work attempts an automatic NL\,$\rightarrow$\,CSP\# pipeline that covers concurrent constructs, shared variables, and refinement assertions.
Establishing such a baseline would therefore require replicating (and fairly re-engineering) the very contribution that PAT-Agent itself introduces, defeating the purpose of comparative evaluation.

Recent verification studies often contrast their approach with program-repair frameworks that iteratively patch counter-examples.  
In our context, however, the combination of prompt engineering and large language models already achieves high requirement-satisfaction rates.
Empirically, incorporating a separate repair pipeline yielded negligible improvements while adding substantial overhead.  
We therefore focus our analysis on synthesis quality rather than post-hoc repair, and leave generalized repair strategies as future work.

\hl{Finally, an important consideration is the generalization ability of the approach. While our implementation is built on CSP\# and PAT, the workflow of planning, code generation, and verification-guided repair is not tied to this particular formalism, and has the potential to generalize to other model checkers~\cite{behrmann2004uppaal,gibson2014fdr3,leuschel2003prob}, and modeling languages such as PEPA and Petri nets~\cite{hillston1993pepa,petri1966communication}.}

\hl{For CSP-style variants (e.g., tock-CSP~\cite{roscoe1998theory}) and for model checkers such as FDR~\cite{gibson2014fdr3}, most prompts and guidance are reusable since the analysis tasks are fundamentally similar, with some effort needed to provide corresponding syntax documentation and example corpora, and process verification feedback. The same workflow can also extend to other languages, with prompt templates and feedback integration adapted to their primitives. }
\section{Threats to Validity}

While \textbf{PAT-Agent} demonstrates promising results in autoformalization, several limitations remain. 

\noindent\textbf{Internal validity.} The effectiveness of PAT-Agent depends on the quality and clarity of the natural language input: ambiguous or incomplete specifications can produce incorrect or underspecified formal models. Moreover, highly entangled control logic or intricate data-dependent behaviors may not decompose cleanly into the semantic-prompt structure we employ, limiting the ability of LLMs to synthesize faithful models. \hl{In addition, like all LLM-based methods, PAT-Agent inherits stochasticity in generation. We mitigate this by fixing LLM versions with timestamps and incorporating multiple repair iterations guided by PAT feedback, but stochasticity remains an inherent limitation.}

\noindent\hl{\textbf{External validity.} Our rule-based repair heuristics are tailored to the violations observed in our datasets, primarily safety and liveness errors. While effective in these settings, they may not generalize to other types of failures, where alternative forms of guidance could be required.}

\section{Related Work}

\noindent\textbf{Autoformalization and Repair.}
Being able to convert the ambiguous natural language descriptions into unambiguous, executable formal specifications is inevitable in ensuring the success of verifications.
Traditionally, this conversion is done manually, with a high requirement for the expertise of the programmer.
The textual understanding capability of LLMs inspires research in autoformalization through LLMs.
For example, research efforts have been dedicated into translating natural language instructions into Linear Temporal Logic (LTL) using LLMs \cite{cosler2023nl2spec, fuggitti2023nl2ltl}.
While LTL is a basic building block of many specification languages, including PAT, it is merely one type of supported assertions in PAT-Agent. The PAT model synthesis task involves constructing and improving models until they pass the given assertions, which is much more complicated.
The other applications of LLMs in formal verification include the synthesis of program specifications \cite{wen2024enchanting, wu2023lemur}, and formal proofs synthesis and repair \cite{first2023baldur, lu2024proof, carrott2024coqpyt}.

On the other hand, existing research on synthesizing system-level models in various specification languages such as Alloy~\cite{hong2025effectiveness} and B-method~\cite{capozucca2025ai} are generally limited in depth. Compared to PAT-Agent, these works target the synthesis of simpler system-level models or fine-grained formulae. In addition, they rely on directly mapping natural language to formal code, which has been demonstrated to be of limited effectiveness~\cite{capozucca2025ai}. In contrast, PAT-Agent offers a more structured and effective solution.

\noindent\textbf{Neuro-symbolic Methods.}
Neuro-symbolic AI combines the pattern-matching power of neural networks with the rigour and compositionality of symbolic reasoning \cite{manhaeve2018deepproblog}.
Recent work attaches large encoders to automated theorem provers: retrieval-augmented LeanDojo boosts success rates on
formal proofs \cite{yang2023leandojo}, and ProofWriter induces natural-language rulebases while recovering proofs \cite{tafjord2021proofwriter}.
Closer to formal verification, FVEL converts real-world code into Isabelle specifications and calls an LLM-guided prover loop \cite{zhang2024fvel}.
Unlike these efforts, we target model checking rather than interactive theorem proving and show how an LLM can populate PAT’s process, after which PAT discharges temporal-logic obligations automatically.

\hlmeta{Another line of work generates verified code from formalized requirements~\cite{cai2025automated}. In contrast, PAT-Agent assumes both a system description and requirements: it first constructs a model from the description and then verifies it against the requirements.}




\section{Conclusion and Future Work}
We presented \textbf{PAT-Agent}, a novel framework that integrates large language models with formal verification to automate the synthesis, validation, and repair of system models from natural language descriptions. 
By leveraging structured semantic prompts and verification-guided feedback, PAT-Agent reliably produces syntactically and semantically correct models, achieving a 100\% verification success rate across 40 tasks.
We plan to expand the semantic prompt library to support more advanced modeling paradigms, such as probabilistic, real-time, and hybrid systems. 
We also aim to generalize the PAT-Agent architecture to other formal methods, including Event-B and Alloy, thereby broadening its applicability. 

\bibliographystyle{plainurl}
\bibliography{reference}
\end{document}